\documentclass[sigconf]{acmart}

\usepackage{acmart-taps}
\usepackage[labelformat=simple]{subcaption}

\usepackage{multirow}
\usepackage{subcaption}
\usepackage{algorithm}
\usepackage{algpseudocode}
\usepackage{amsmath}
\usepackage{xcolor}

\usepackage{calc}
\usepackage{fancybox,framed}
\usepackage[ruled,lined,linesnumbered,vlined,algo2e]{algorithm2e}

\newenvironment{boxedtext}
{\noindent\begin{Sbox}\begin{minipage}{\linewidth-7.5\fboxrule-2\fboxsep-1pt}}
{\end{minipage}\end{Sbox}\doublebox{\TheSbox}}

\usepackage{array}
\usepackage{appendix}
\usepackage{multirow}
\usepackage{enumitem}
\usepackage[utf8]{inputenc}
\usepackage{pifont}
\usepackage{amsthm}

\usepackage{xspace}
\usepackage{makecell}

\newcommand{\tool}{CCScanner\xspace}

\newcommand{\centris}{\textsc{CENTRIS}\xspace}
\newcommand{\fossid}{\textsc{FOSSID}\xspace}
\newcommand{\owasp}{\textsc{OWASP}\xspace}
\newcommand{\sonatype}{Sonatype\xspace}
\newcommand{\ccpp}{C/C++\xspace}
\newcommand{\saveSpaceFig}{\vspace{-6pt}}

\newcommand{\Fig}{Figure\xspace}
\newcommand{\Tab}{Table\xspace}
\newcommand{\Sec}{Section\xspace}

\newcommand{\phaseone}{Install\xspace}
\newcommand{\phasetwo}{Build\xspace}

\newcommand{\ccrepodataset}{\ccpp Repo Dataset\xspace}
\newcommand{\tpldataset}{TPL Dataset\xspace}

\copyrightyear{2022} 
\acmYear{2022} 
\setcopyright{acmlicensed}
\acmConference[ASE '22]{37th IEEE/ACM International Conference on Automated Software Engineering}{October 10--14, 2022}{Rochester, MI, USA}
\acmBooktitle{37th IEEE/ACM International Conference on Automated Software Engineering (ASE '22), October 10--14, 2022, Rochester, MI, USA}
\acmPrice{15.00}
\acmDOI{10.1145/3551349.3560432}
\acmISBN{978-1-4503-9475-8/22/10}

\begin{document}
\title{Towards Understanding Third-party Library Dependency in \ccpp Ecosystem}




\author{Wei Tang\textsuperscript{$^1$}, Zhengzi Xu\textsuperscript{$^2$$\ast$}, Chengwei Liu\textsuperscript{$^2$}, Jiahui Wu\textsuperscript{$^2$}, Shouguo Yang\textsuperscript{$^3$}, Yi Li\textsuperscript{$^2$}\\ Ping Luo\textsuperscript{$^1$} and Yang Liu\textsuperscript{$^2$}}

\thanks{$\ast$ Corresponding author}
\affiliation{%
  \institution{$^1$ School of Software, Tsinghua University \country{Beijing, China}}
  \institution{$^2$ School of Computer Science and Engineering, Nanyang Technological University \country{Singapore}}
  \institution{$^3$ Institute of Information Engineering, Chinese Academy of Sciences \country{Beijing, China}}
}
\email{tang-w17@mails.tsinghua.edu.cn,{zhengzi.xu, yi_li, yangliu}@ntu.edu.sg}
\email{{chengwei001, jiahui004}@e.ntu.edu.sg, yangshouguo@iie.ac.cn}

\renewcommand{\shortauthors}{Tang et al.}


\begin{abstract}
Third-party libraries (TPLs) are frequently reused in software to reduce development cost and the time to market. However, external library dependencies may introduce vulnerabilities into host applications. The issue of library dependency has received considerable critical attention. Many package managers, such as Maven, Pip, and NPM, are proposed to manage TPLs. Moreover, a significant amount of effort has been put into studying dependencies in language ecosystems like Java, Python, and JavaScript except \ccpp. Due to the lack of a unified package manager for \ccpp, existing research has only few understanding of TPL dependencies in the \ccpp ecosystem, especially at large scale.

Towards understanding TPL dependencies in the \ccpp ecosystem, we collect existing TPL databases, package management tools, and dependency detection tools, summarize the dependency patterns of \ccpp projects, and construct a comprehensive and precise \ccpp dependency detector. Using our detector, we extract dependencies from a large-scale database containing 24K \ccpp repositories from GitHub. Based on the extracted dependencies, we provide the results and findings of an empirical study, which aims at understanding the characteristics of the TPL dependencies. We further discuss the implications to manage dependency for \ccpp and the future research directions for software engineering researchers and developers in fields of library development, software composition analysis, and \ccpp package manager.
\end{abstract}



\maketitle


\section{Introduction}\label{sec:intro}

Reusing third-party libraries (TPLs) as dependencies has been widely adopted as a common practice to save time and manpower in software development. However, TPL reuse could also constantly expose downstream projects to potential risks of being attacked via vulnerabilities from dependencies.
Therefore, acknowledging, monitoring, and managing TPLs that are introduced as dependencies, and further mitigating potential vulnerability threats, are critical demands in modern software development.

To facilitate the reuse of TPLs, many practical package managers have been proposed to manage dependencies, especially for younger languages. For example, NPM, Pip, and Maven are powerful package managers for Node.js, Python, and Java, respectively. However, for \ccpp, which is an ancient programming language, although millions of components are available and various solutions for reusing TPLs have been proposed and adopted, there is still no unified package manager~\cite{miranda2018use, wu2015developers} that can manage the dependencies of \ccpp projects well, which significantly complicates the dependency management of \ccpp projects and hinders the process of embracing agile DevOps~\cite{DevOpsWi24:online} in \ccpp.


Unlike other languages in that researchers have gained a lot of in-depth insights,
for \ccpp, people might have only few understanding of the fundamental aspects. 
For example, questions, such as how developers introduce \ccpp TPLs into their projects and what the TPL data scope is, are not comprehensively studied.
Due to the lack of a standard package format and unified package manager, there are no available methods and tools to extract dependencies against large-scale \ccpp repositories. Consequently, people have no insights on TPL dependencies at large scale, and even no awareness of TPL data scope in the \ccpp ecosystem. Insights on TPL dependency landscape are important and beneficial for numerous areas. For example, OSSPolice~\cite{duan2017identifying} detects \ccpp TPL dependencies in Android applications. It relies on a local TPL database for detection. With unawareness of TPL data scope, OSSPolice collects GitHub repositories as TPLs that contain a large number of internal code clones~\cite{lopes2017dejavu, kalliamvakou2016depth}, further resulting in false positives. To bridge this knowledge gap, we seek to conduct a large-scale empirical study on dependencies in \ccpp ecosystem.

For an empirical study on dependencies to be successful and insightful, it is crucial to build a \textbf{comprehensive}, \textbf{precise}, and \textbf{efficient} detector for TPL reuse detection in large-scale repositories. A large number of existing researches~\cite{woo2021centris, lopes2017dejavu, fang2020functional, jiang2007deckard} conducted code clone detection to extract TPL dependencies in \ccpp repositories. However, not all dependencies are introduced by code clone. Miranda et al.~\cite{miranda2018use} conducted a survey with a questionnaire involving 343 \ccpp developers. They report that there are multiple methods that developers prefer to handle dependencies in \ccpp projects. The top favorite methods to add dependencies are system package manager, header-only libraries, and git submodules. Only 10\% of developers put the source code of libraries in host repositories, which suggests that code clone detection methods might only have a limited ability to detect dependencies. 
\begin{figure*}[t]
    \centering
    \includegraphics[width=0.8\textwidth]{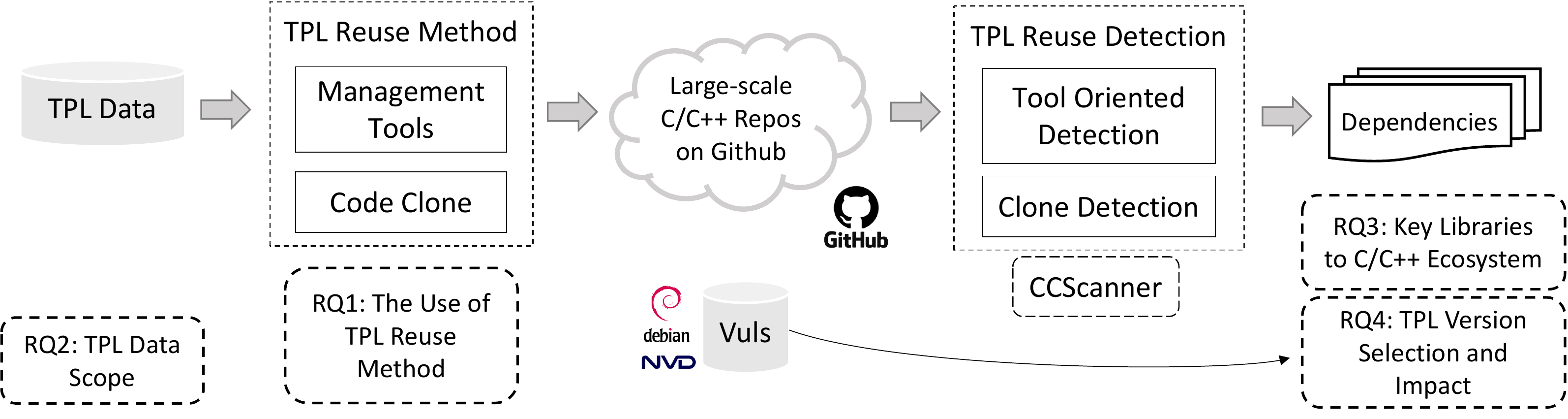}
    \saveSpaceFig
    \caption{Overview of our work}
    \saveSpaceFig
    \label{fig:overview}
\end{figure*}

In this paper, to unify the dependency management and detect dependencies of \ccpp properly, we first undertake a detailed investigation on how TPL dependencies are handled from the original published TPL data to the final dependent application and summarize the lifecycle for dependencies. Based on the dependency lifecycle, we propose a dependency detector that analyzes every step where dependencies might be introduced in the lifecycle and parses all possible TPL reuse methods which create dependencies. As shown in \Fig~\ref{fig:overview}, we divide TPL reuse methods into two categories depending on whether tools are used or not: \textbf{1)} package management tools and \textbf{2)} code clone. For both reuse methods, we design two corresponding modules for the TPL detection. We build a dependency detector, \tool, with the ability to detect dependencies introduced through 21 management tools. Moreover, its clone detection module integrates a state-of-the-art tool~\cite{woo2021centris} for identifying \ccpp TPL reuse. Our experiments show that \tool can precisely and efficiently extract dependencies from large-scale repositories with the highest recall of 80.1\%.

For a large-scale empirical study, we download a collection of 24K GitHub \ccpp repositories that own more than 100 stargazers and scan all repositories to extract dependencies using \tool. Totally, we obtain over 150K TPL dependencies from 24K repositories, based on that, we conduct an empirical study on dependencies in the \ccpp ecosystem. Through our empirical study, we conclude some key findings as follows. \textbf{1)} Developers prefer to add dependencies unintentionally in build scripts (over 70\% of dependencies), not explicitly managing dependency installation. This convention significantly constrains the effectiveness of existing dependency detection tools. \textbf{2)} TPL data fragmentation exists in the \ccpp ecosystem. Libraries are inconsistent between databases, which may threaten the effectiveness of TPL detection and vulnerability reporting. \textbf{3)} System libraries on OS are the most important libraries for the \ccpp ecosystem. A group of 10 popular system libraries has an impact on the entire ecosystem. However, system libraries are always neglected in TPL management and detection. \textbf{4)} Management tools that rely on system libraries will introduce vulnerabilities from OS environments (22\% of repositories). Half of version specifications of vulnerable libraries use vulnerable versions, that require developers to update the dependencies manually. Based on our findings, we present some recommendations for practitioners in related areas and future directions.

In summary, our contributions are as follows:
\begin{itemize}[leftmargin=4mm]
\item We describe the lifecycle of \ccpp dependencies including mechanisms to import dependencies, dependency management tools, and TPL database for \ccpp dependencies.
\item We propose a comprehensive and precise \ccpp dependency detector, \tool, based on the dependency lifecycle. The evaluation result demonstrates that \tool can achieve a precision of 86\% and a recall of 80.1\%. Besides, \tool is capable of efficiently scanning large-scale \ccpp repositories for empirical studies.
\item We conduct a large-scale empirical study on \ccpp dependencies and discuss the research questions about the method to import dependencies, TPL data scope for \ccpp dependencies, key library in the \ccpp ecosystem, and TPL version constraints. We provide useful insights for helping improve the capability of package management for \ccpp. We make our code and data available\footnote{https://github.com/lkpsg/ccscanner}.
\end{itemize}



\section{Dependency Lifecycle Overview}\label{sec:dependency_cycle}
Dependency lifecycle is a concept that describes the process of TPL from the database where it is stored to applications where it is reused. Understanding how dependencies are processed in their lifecycles would guide us to build an effective dependency detector.

Referring to \verb|Maven| for Java, we present how a TPL is processed in its lifecycle according to TPL location. \verb|Maven| downloads all TPLs from the Maven central repository and put them under a local directory, like \verb|~/.m2|. Then, dependencies are packaged into the dependent project. We summarize that general TPL management mainly consists of two phases, \phaseone and \phasetwo, that are separated by TPL location. The \phaseone phase is responsible for installing third-party packages under local directories and the \phasetwo phase compiles and links libraries into final artifacts.

\begin{figure}[t]
    \centering
    \includegraphics[width=0.88\columnwidth]{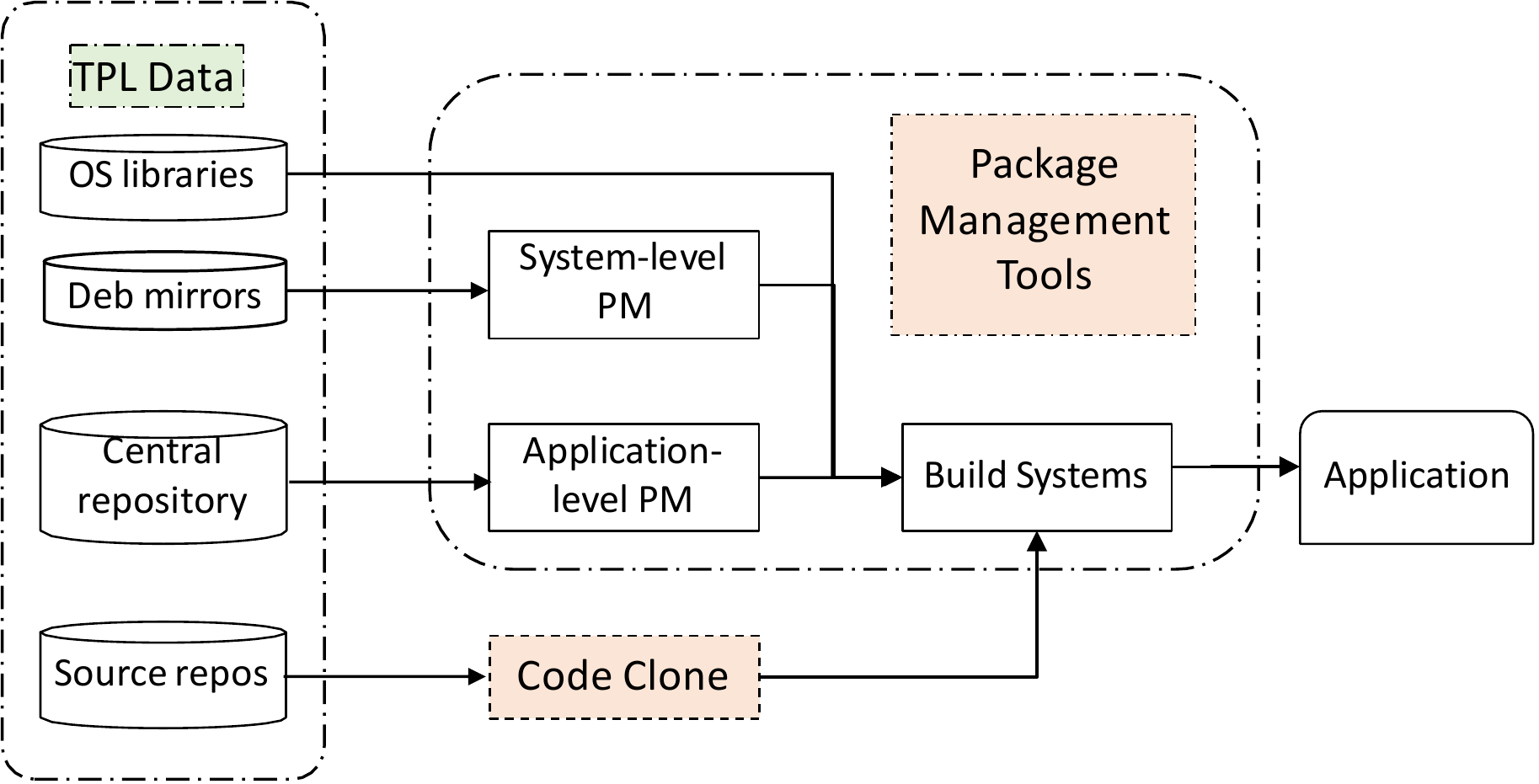}
    \saveSpaceFig
    \caption{\ccpp TPL dependency lifecycle and toolchains. PM: pacakge manager}
    \saveSpaceFig
    \label{fig:dependency_lifecycle}
\end{figure}

Unlike other languages, there is no unified and language-specific package management systems for \ccpp, which means it is impossible to collect complete datasets from a single data source.
Moreover, as far as we have reviewed, there is no relevant literature on what the library reuse mechanisms are and how to collect the aforementioned datasets. Therefore, it is the foundation and prerequisite for further investigation to figure out the TPL lifecycle when \ccpp projects reuse TPLs. With extensive research on previous studies \cite{miranda2018use, wu2015developers} and related real-world projects, we have summarized the \ccpp dependency lifecycle including TPL databases and TPL reuse methods as shown in \Fig~\ref{fig:dependency_lifecycle}.

There are two types of colorful boxes, TPL data and TPL reuse methods. TPL data consists of data sources from which the reusable libraries can be retrieved. It includes default installed libraries on OS, mirrors of system-level package managers, central repositories of application-level package managers, and source repositories that are hosted on platforms such as GitHub or official websites. TPL reuse methods include package management tools and code clone, that introduce reused TPL in the two phases, \phaseone and \phasetwo. Package management tools contain system-level and application-level package managers for installing dependencies in \phaseone phase and build systems for \phasetwo phase. Different TPL databases and reuse methods work together as toolchains to complete the dependency management task during the entire lifecycle, and the toolchains can be classified into three types as follows:


\begin{itemize}[leftmargin=4mm]
    \item \textbf{System library toolchain:} This toolchain retrieves TPLs from default installed libraries on operating systems (OS) and the mirrors of system-level package managers, such as Debian mirrors~\cite{debianmirror}. System-level package managers like \verb|APT|~\cite{apt} install libraries under system paths globally on OS. Build systems search reused libraries under system paths to compile and link~\cite{find_library}. It is clear that system-level package managers are system-specific, not language-specific. Therefore, different systems have different toolchains.
    For example, \verb|APT| on Linux, \verb|Homebrew|~\cite{homebrew} on MacOS and \verb|winget|~\cite{winget} on Windows are toolchains for mainstream OS.
    \item \textbf{Application-level package manager toolchain:} \ccpp has long been lacking a popular language-specific application-level package manager. \ccpp application-level managers are used to download TPLs from their central repositories for \ccpp. Different from system-level package managers, this kind of package manager is language-specific like \verb|Maven| and installed at the application level. They reside within a directory that is not maintained by the system environment. 
    Generally, an official central repository of TPLs is built and maintained along with the application-level package manager. One of the famous \ccpp package managers is \verb|Conan|~\cite{url:conan}.
    Package managers need to integrate with build systems to specify the library directory path. Subsequently, build systems could link the libraries and compile the whole project.
    \item\textbf{Code clone toolchain:} In this toolchain, libraries are reused by code clones that are not downloaded from specified TPL databases. The code clone method directly puts the source code of libraries into host repositories. It is a primary and quick way to import a dependency. Some techniques like \verb|git submodules| help to manage the TPL versions and fetch source code automatically. The access to source code allows developers to modify the code of libraries to improve security, remove unnecessary code, or add new functionalities. However, reusing libraries through source code means that developers need to compile everything from source code on their own.
\end{itemize}

The differences between the three types of toolchains are the data source in the \phaseone phase. An end-to-end toolchain completes the management functionality for the whole dependency lifecycle like \verb|Maven|. However, not all repositories utilize the complete toolchain. For example, \verb|Ifopt|~\cite{url:ifopt} only contains build scripts without the use of package managers. It provides a \verb|README.md| to instruct developers to install required dependencies manually through \verb|apt-get| command. We consider this reuse method as an incomplete system library toolchain. It is possible to miss any step in the toolchain for a \ccpp project since there is no unified end-to-end package manager.

In addition to these toolchains, SBOM~\cite{sbom} (software bill of materials) files can be used to describe dependency relationships and specify the source of libraries. The are some standard SBOM formats, such as \verb|SPDX|~\cite{url:spdx} and \verb|CycloneDX|~\cite{url:cyclonedx}. Large-scale development teams may define their own formats like \verb|Firefox|~\cite{url:updatebot} and \verb|Chromium|~\cite{url:chromiundocs}. However, less than 10 repositories in our 24K \ccpp repositories adopt SBOM to manage dependencies, and SBOM files only provide the description of dependencies. Therefore, we would disregard these SBOM rules in this paper.

\section{Dependency detection}
\subsection{\tool}
\begin{figure}[t]
    \centering
    \includegraphics[width=0.7\columnwidth]{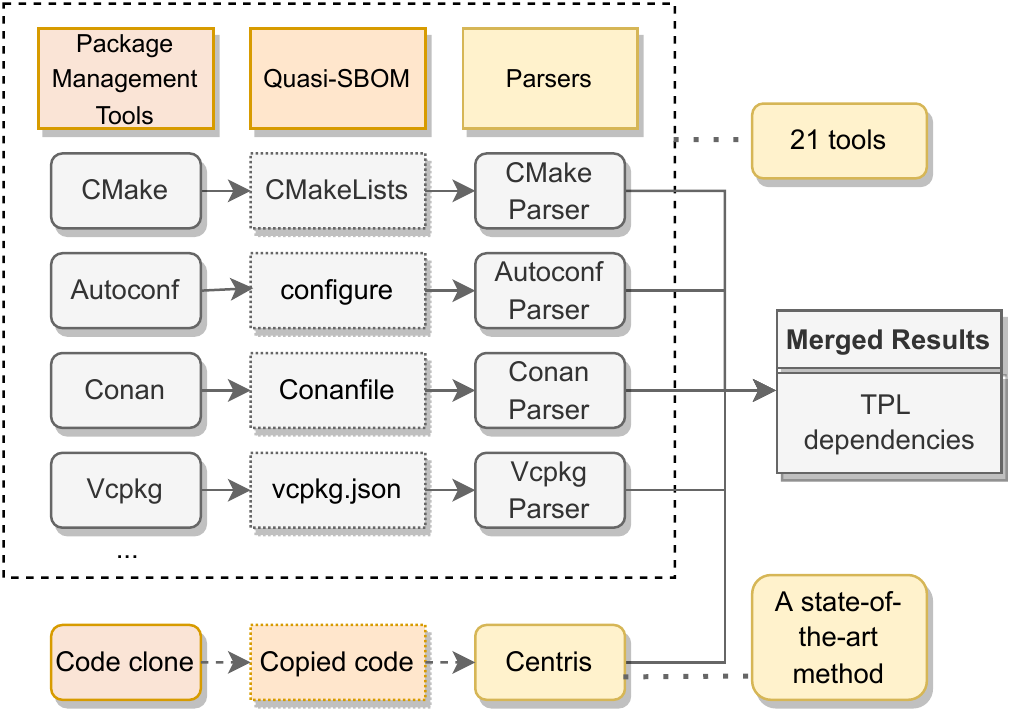}
    \saveSpaceFig
    \caption{CCScanner}
    \saveSpaceFig
    \label{fig:ccscanner}
\end{figure}
Dependencies in Java projects could be easily extracted by parsing SBOM files like \verb|pom.xml| in \verb|Maven|. However, it is difficult for \ccpp since there is no central SBOM to store all package information. As mentioned in \Sec~\ref{sec:dependency_cycle}, three types of toolchains work as \verb|Maven| to retrieve TPLs, compile and package artifacts. Numerous available methods and tools can be selected at every step in the toolchain. Depending on whether tools are used, we divide elements that can be used to extract dependencies into two categories, \textbf{quasi-SBOM files} in package management tools and \textbf{copied code} with code clone method as shown in \Fig~\ref{fig:ccscanner}.

\begin{figure}[t]
    \centering
    \includegraphics[width=0.6\columnwidth]{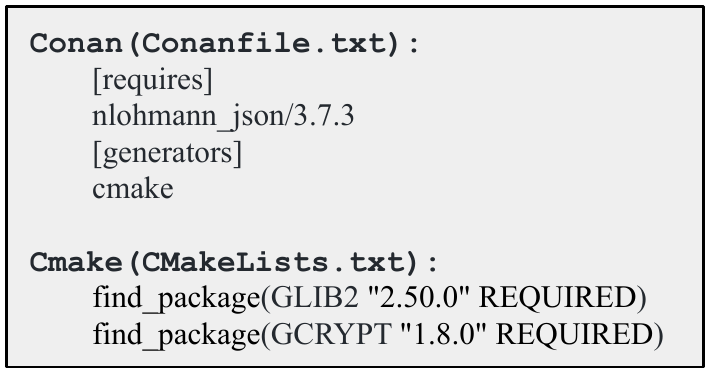}
    \saveSpaceFig
    \caption{Examples of quasi-SBOM files for Conan\protect~\cite{url:conan_project} and CMake\protect~\cite{url:cmake_project}.}
    \saveSpaceFig
    \label{fig:examples_of_sbom}
\end{figure}

\noindent\textbf{Quasi-SBOM files} Package management tools always need configuration files or specific statements to specify what libraries will be introduced. Therefore, these files can be recognized as a kind of quasi-SBOM files that describe required dependencies. Examples of quasi-SBOM and statements for Conan and CMake are shown in \Fig~\ref{fig:examples_of_sbom}. When using \verb|Conan| to install libraries, \verb|Conanfile.txt| file is written to describe package information. Dependencies can be extracted by parsing the \verb|[requires]| field in \verb|Conanfile.txt|. \verb|CMake| finds external packages using \verb|find_package| method. Libraries would be specified in \verb|CMakeLists.txt| as arguments of \verb|find_package|. A corresponding analyzer could be built to parse quasi-SBOM files based on the syntax of the package management tool.

After extensive research, we have summarized all tools that can be utilized in toolchains. Tools can be divided to package managers and build systems as shown in \Tab~\ref{tab:elements}. In total, we have found 21 package management tools that are used by \ccpp developers. For system-level package managers, we select the Debian package manager since it is widely used in the \ccpp ecosystem. Other system package managers, such as \verb|RPM|~\cite{url:rpm} and \verb|Homebrew|~\cite{homebrew}, are similar and not considered in this paper. The Name column in \Tab~\ref{tab:elements} shows the names of corresponding quasi-SBOM files we parse for each tool. More details can be seen in our public code repo of \tool~\cite{ccscanner}.

\noindent\textbf{Copied code} The code clone method copies and pastes reused code into the host repository and does not utilize any management tool. Consequently, there are no available quasi-SBOM files that describe the library information. In this case, the copied code can be used to generate features for code clone detection systems. TPL dependencies would be detected by matching similar code against a local TPL code database. For TPLs imported by code clone method, we utilize \centris~\cite{woo2021centris}, a state-of-the-art system, to detect \ccpp TPL dependencies.


\aptLtoX[graphic=no,type=env]{%
\begin{table}[t]
  \caption{Elements used in \tool.}
  \label{tab:elements}
  \begin{tabular}{lcc}
    \toprule
        Type & Tools & SBOM/Method\\
    \midrule
        \multirow{10}{*}{\makecell{Package\\Managers}} & Deb & Control\\
        &Conan & conanfile.*, conaninfo.txt\\
        &Vcpkg & vcpkg.json\\
        &Clib  & package.json, clib.json\\
        &CPM   & CMakeLists.txt\\
        &Buckaroo & buckaroo.toml\\
        &Dds   & package.json5\\
        &Hunter& CMakeLists.txt\\
        &Cppget& manifest\\
        &Xrepo & xmake.lua\\
        &Gitsubmodule & .gitmodules\\
        &Pkg-config & *.pc\\
    \midrule
        \multirow{9}{*}{\makecell{Build\\Systems}} & Make & Makefile\\
        &CMake & CMakeLists.txt, *.cmake\\
        &Autoconf & configure, configure.*\\
        &Bazel & bazel.build, BUILD\\
        &Meson & meson.build\\
        &MSBuild & *.vcxproj, *.vbproj, *.props\\
        &Xmake & xmake.lua\\
        &Build2 & manifest\\
        &Buck & BUCK\\
    \midrule
        Code clone & Centris & Code clone detection\\
    \bottomrule
\end{tabular}
\end{table}
}{%
\begin{table}[t]
  \renewcommand\arraystretch{1}
  \caption{Elements used in \tool.}
  \small
  \label{tab:elements}
  \scalebox{0.86}{
  \begin{tabular}{lcc}
    \toprule
        Type & Tools & SBOM/Method\\
    \midrule
        \multirow{10}{*}{\makecell{Package\\Managers}} & Deb & Control\\
        &Conan & conanfile.*, conaninfo.txt\\
        &Vcpkg & vcpkg.json\\
        &Clib  & package.json, clib.json\\
        &CPM   & CMakeLists.txt\\
        &Buckaroo & buckaroo.toml\\
        &Dds   & package.json5\\
        &Hunter& CMakeLists.txt\\
        &Cppget& manifest\\
        &Xrepo & xmake.lua\\
        &Gitsubmodule & .gitmodules\\
        &Pkg-config & *.pc\\
    \midrule
        \multirow{9}{*}{\makecell{Build\\Systems}} & Make & Makefile\\
        &CMake & CMakeLists.txt, *.cmake\\
        &Autoconf & configure, configure.*\\
        &Bazel & bazel.build, BUILD\\
        &Meson & meson.build\\
        &MSBuild & *.vcxproj, *.vbproj, *.props\\
        &Xmake & xmake.lua\\
        &Build2 & manifest\\
        &Buck & BUCK\\
    \midrule
        Code clone & Centris & Code clone detection\\
    \bottomrule
\end{tabular}
}
\saveSpaceFig
\end{table}
}

Based on all management tools and TPL reuse methods, we build a comprehensive \ccpp dependency detector, \tool. It has the capability to parse the quasi-SBOM files of 21 package management tools and integrates \centris to detect cloned TPLs. To the best of our knowledge, \tool has the best capability to detect \ccpp dependencies compared to existing techniques. Each existing tool is only capable of dealing with one or two elements in \Tab~\ref{tab:elements}. For example, \centris can only detect dependencies with cloned code. The existing state-of-the-art SBOM parser tool, \owasp~\cite{url:denpcheck} can only deal with \verb|CMake| and \verb|Autoconf| for \ccpp.

\subsection{Evaluation of \tool}
In this section, we construct the ground truth for evaluation and present the accuracy evaluation results of \tool to prove that it is effective to scan large-scale databases for empirical study. We compare \tool with related works including two clone detection tools, \centris (a state-of-the-art tool for cloned \ccpp TPL detection) and \fossid (a famous commercial software composition analysis tool), and two state-of-the-art SBOM scanners, \owasp Dependency-Check and \sonatype. We present our findings on \ccpp dependency detection and discuss the effectiveness of existing tools.

\subsubsection{\textbf{Ground truth Construction}}
Since there is no public test data for \ccpp dependency detection, we need to establish the ground truth dataset containing a labeled mapping of \ccpp repositories and reused TPL dependencies. We adopt 15 real-world projects that cover all categories, such as audio processing, database management, and Internet connection, from the dataset of previous work~\cite{miranda2018use} that analyzed the use of package managers by \ccpp developers. 
To label the dependency relationships, we manually check all files including source code files and non-source code files in projects to find TPL dependencies. After manually checking that took over 100 hours, we obtained 589 TPL dependencies as the ground truth.

\aptLtoX[graphic=no,type=env]{%
\begin{table}[t]
  \caption{Performance comparison of SCA tools. P: Precision, R1: Recall on the full ground truth, R2: Recall on dependencies that are supported to detect. F1: 2*P*R1/(P+R1).}
  \label{tab:sca_performance}
        \begin{tabular}{lcccc}
        \toprule
        Method & P (\%) & R1 (\%)& R2 & F1\\\midrule
        \centris & 33.6 & 5.3 & 22.4 & 0.09\\
        \fossid & 12.7 & 2.1 & 12.5 & 0.04\\
        \midrule
        \owasp & 93.9 & 27.2 & 83.3 & 0.42\\
        \sonatype & / & 0 & 0 & /\\
        \midrule
        \tool & 86.0 & 80.1& 80.8 & \textbf{0.83}\\
        \bottomrule
        \end{tabular}
\end{table}
}{%
\begin{table}[t]
  \caption{Performance comparison of SCA tools. P: Precision, R1: Recall on the full ground truth, R2: Recall on dependencies that are supported to detect. F1: 2*P*R1/(P+R1).}
  \label{tab:sca_performance}
  \small
    \scalebox{1}{
        \begin{tabular}{lcccc}
        \toprule
        Method & P (\%) & R1 (\%)& R2 & F1\\\midrule
        \centris & 33.6 & 5.3 & 22.4 & 0.09\\
        \fossid & 12.7 & 2.1 & 12.5 & 0.04\\
        \midrule
        \owasp & 93.9 & 27.2 & 83.3 & 0.42\\
        \sonatype & / & 0 & 0 & /\\
        \midrule
        \tool & 86.0 & 80.1& 80.8 & \textbf{0.83}\\
        \bottomrule
        \end{tabular}
    }
  \saveSpaceFig
\end{table}
}

\subsubsection{\textbf{Accuracy analysis}}\label{accuracy}
We evaluate \tool and baselines on our ground truth and calculate the precision and recall. There are two kinds of recall rates, R1 and R2. R1 refers to the recall rate against full ground truth. However, each baseline is designed for only one or two elements in \Tab~\ref{tab:elements}. We calculate R2 for each baseline based on the dependencies that the baseline supports to detect. From the results in \Tab~\ref{tab:sca_performance}, we can see that \tool outperforms all baselines in terms of F1 score.

The R1 and R2 of \tool are nearly identical, which means the detection capability of \tool can cover most elements used in ground truth. More than 90\% of false results of \tool are caused by the code clone module provided by \centris since we integrate unchanged \centris for the detection. Without \centris,
\tool achieves both accuracy and recall rate of 97\%. Sometimes, the SBOM parser in \tool makes mistakes in processing sophisticated syntax, such as complex variable replacement. Even though, our results prove that the SBOM parser module of \tool significantly outperforms \owasp Dependency-Check and \sonatype.

We inspected the false results reported by \centris module and found that even though the \centris module reports false original reused libraries, there truly exist code reuse and TPL dependency for each reported positive case since \centris applies identical hashes to match reused code. Besides, we have found that many dependencies are reused by code clone method, but they only account for a small part of the total dependencies. In our ground truth, code clone introduces 24.4\% of dependencies, that are not managed by other tools. As a result, \centris can still be integrated with the SBOM module for large-scale empirical study. Reasons for false results of \centris are illustrated in \Sec~\ref{existing_sca_tools}.

\aptLtoX[graphic=no,type=env]{%
\begin{framed}
Our experiments show that \tool is comprehensive and precise for \ccpp TPL dependency detection. We use it to scan a large-scale database containing 24K \ccpp repositories and conduct an empirical study on dependencies in \ccpp ecosystem.
\end{framed}
}{%
\vskip 2mm
\begin{boxedtext}
Our experiments show that \tool is comprehensive and precise for \ccpp TPL dependency detection. We use it to scan a large-scale database containing 24K \ccpp repositories and conduct an empirical study on dependencies in \ccpp ecosystem.
\end{boxedtext}
}



\subsubsection{\textbf{The effectiveness of existing SCA tools}}\label{existing_sca_tools}
Based on our experiments, we discuss the effectiveness of existing SCA tools for better understandings of \ccpp dependency detection.

SBOM parser-based tools generally have high precision, since the SBOM formats are strictly defined by package managers. However, both state-of-the-art SBOM scanners, \owasp and \sonatype, have low recall rates. \sonatype does not report any dependency, because it only supports scanning SBOM files in Conan. There is no unified package manager or SBOM format in \ccpp ecosystem and multiple tools are used in the dependency lifecycle. The comprehensiveness of SBOM parser is the most important factor to the recall rate. Therefore, it is critical to understand the use of package managers. It would be discussed in our empirical study for RQ1 in \Sec~\ref{sec:rq1}.

For code clone detection-based tools, both \centris and \fossid perform poorly, especially on the recall. Such tools search similar or identical context hashes against a local TPL database. The reason for false positives is that duplication exists across different libraries. \fossid has a lower precision since it adds non-source code files as features, unlike \centris which only uses source code files. Duplicated context of license or description would cause false positive results. The reason for false negatives is the lack of reused libraries in TPL database. In general, the quality of TPL database is a determining factor for the effectiveness of the tool. How to build a TPL database is a critical research question for the field of dependency detection. It would be discussed for RQ2 in \Sec~\ref{sec:data scope}.




\section{Experimental Design}\label{sec:exp}
\subsection{Dataset}
We collect two datasets for large-scale empirical study including a large-scale database of popular \ccpp repositories and a TPL database.

\noindent\textbf{\ccrepodataset} To investigate dependencies in the \ccpp ecosystem, we extract dependencies from a large-scale collection of \ccpp repositories that are popular and receive recognition from the \ccpp community. \ccpp GitHub repositories with more than 100 stargazers are often used in previous works~\cite{duan2017identifying, woo2021centris} and the number of repositories increased to 24,001 in Mar 2022, which is adequate for the study. Therefore, we pull all 24K git repositories containing 3.5 billion lines of code as detection targets for the analysis on \ccpp ecosystem.

\noindent\textbf{\tpldataset} As aforementioned, TPL databases 
can be divided into three categories: system libraries including OS libraries and mirrors of system-level package managers, central repositories of application-level package managers, and source repositories on the Internet. 
For OS libraries, We scan the OS environment and collect default installed libraries under the standard directory such as "/usr/lib" to form an OS library database.
In addition to them, the metadata of all libraries in the Debian mirror is crawled. We combine OS libraries and Debian mirror together to build the system library database.
For the central repository of application-level package managers, we collected and merged repositories of 7 package managers including Conan, Vcpkg, Clib, Cppget (Build2), Hunter, Buckaroo and Xrepo (Xmake). The merged repository contains 3380 reusable libraries. 
For source repositories, we select the database of the \centris module as the third category of TPL database.


\section{Empirical Study}\label{sec:eva}
In the experiments, we use \tool to conduct a large-scale \ccpp dependency analysis against 24K GitHub repositories with more than 100 stars. In total, we obtain over 150K TPL dependencies from 24K repositories with the latest versions to study the TPL dependencies in the \ccpp ecosystem. Since the lack of understanding of many fundamental aspects, we first look into TPL reuse method and TPL data in the \ccpp ecosystem as shown in \Fig~\ref{fig:overview}. Then, we study key libraries and version constraints caused by large-scale reuses. In summary, we try to investigate the following research questions (RQs):
\begin{itemize}[leftmargin=4mm]
\item RQ1: The use of TPL reuse methods to handle dependencies.
\item RQ2: TPL data scope for \ccpp.
\item RQ3: Key libraries in the \ccpp ecosystem.
\item RQ4: TPL version selection and the impact of vulnerable versions.
\end{itemize}

\subsection{RQ1: The Use of TPL Reuse Method}\label{sec:rq1}
In this section, we will investigate the use of TPL reuse methods in dependency lifecyle. Even though existing studies~\cite{miranda2018use} have discussed what TPL reuse methods are adopted to add dependencies in \ccpp projects, insufficient empirical evidence on dependencies and details of methods were provided. Therefore, we aim to re-investigate this question via large-scale analysis. As described in \Sec~\ref{sec:dependency_cycle}, multiple methods and tools are combined to form a toolchain in two phases. We investigate each step in the dependency lifecycle to answer three main questions, (1) which phase is preferred to introduce TPL dependencies? (2) what tools are reused more often? (3) how are the tools combined as toolchains? The following three subsections answer these questions respectively.

\subsubsection{\textbf{Preferred Phase}}
We first count how many dependencies are handled and how many repositories are involved in each phase. Statistics indicate that 71.5\% of dependencies are extracted from \phasetwo phase and 80.3\% of repositories adopt build systems. However, only 47.5\% of repositories adopt methods and tools in \phaseone phase that introduces 37.5\% of dependencies. It proves that TPL management is separated in the dependency lifecycle. \phasetwo phase is complicated for \ccpp projects to generate binaries and support enormous platforms, therefore the build scripts are provided along with source code in TPLs. However, \ccpp projects are not required to provide management scripts to install dependencies. Sometimes, developers write customized documents to guide system-level package managers to install dependencies. However, it is not an explicit and formal method for TPL management and detection. It can be seen that \ccpp projects do not perform well on explicitly managing dependency installation using tools in \phaseone phase. However, most of the existing SCA tools focus on \phaseone phase including code clone detection and SBOM scan. It means they would neglect dependencies added in \phasetwo phase that occupy a large portion. Only 9\% of dependencies are handled in both two phases from \phaseone to \phasetwo, which means a lot of manual efforts are necessary to handle the whole lifecycle for \ccpp dependencies.

\aptLtoX[graphic=no,type=env]{%
\begin{framed}
Finding-1: \ding{172} Developers prefer adding dependencies unintentionally in \phasetwo phase (over 70\% of dependencies), not explicitly managing the installation of dependencies in \phaseone phase. This convention significantly constrains the effectiveness of existing dependency detection tools. \ding{173} Due to the separation of dependency lifecycle for \ccpp, developers usually do not maintain complete toolchains for dependencies. Only 9\% of dependencies are handled by complete toolchains. A lot of manual efforts are required for TPL reuse.
\end{framed}
}{%
\vskip 2mm
\begin{boxedtext}
Finding-1: \ding{172} Developers prefer adding dependencies unintentionally in \phasetwo phase (over 70\% of dependencies), not explicitly managing the installation of dependencies in \phaseone phase. This convention significantly constrains the effectiveness of existing dependency detection tools. \ding{173} Due to the separation of dependency lifecycle for \ccpp, developers usually do not maintain complete toolchains for dependencies. Only 9\% of dependencies are handled by complete toolchains. A lot of manual efforts are required for TPL reuse.
\end{boxedtext}
}





\subsubsection{\textbf{Usage of Tools}}
For the usage of each tool, we count the proportion of dependencies that are managed by the tool and the proportion of repositories that adopt it. Results are shown in \Tab~\ref{tab:popularity_pm}. The code clone method is excluded since it is a general method rather than a specific tool. Two package managers, Cppget and Xrepo are excluded since they are a part of Build2 and XMake respectively. Dds is not in the table, because it is not used by any repository in \ccrepodataset.

Shown in \Tab~\ref{tab:popularity_pm},  the most popular tool in \phaseone phase is \verb|Gitsubmodule|, that is actually a tool for automatic code clone. Besides, we notice that 32.6\% of repositories adopt code clone method and introduce 15.77\% of dependencies. Code clone is more popular than tools in \phaseone phase. This is a disappointing fact that \ccpp developers prefer to explicitly specify resources in \phaseone phase by code clone, which will hinder the package management process.


\aptLtoX[graphic=no,type=env]{%
\begin{table}[t]
  \caption{The use of package management tools.}
  \label{tab:popularity_pm}
  \begin{tabular}{lcclcc}
    \toprule
        Install & Dep.(\%) & Repo.(\%) & Build & Dep(\%) & Repo(\%)\\
    \midrule
        Gitsubmod & 4.94 & 16.79& Make Only & 2.27 & 25.00 \\
        Deb & 9.32 & 5.90 & Cmake & 51.33 & 39.11 \\
        Pkg-config & 6.94 & 1.83 & MSBuild & 5.34 & 16.86 \\
        Conan & 0.45 & 0.99 & Autoconf & 12.3 & 15.27 \\
        Vcpkg & 0.56 & 0.39 & Meson & 3.0 & 2.11 \\
        Hunter & 0.18 & 0.2 & Bazel & 0.81 &1.37\\
        Clib & 0.05 & 0.12 & Buck & 0.0 & 0.27 \\
        Cpm & 0.07 & 0.1 & Xmake & 0.08 & 0.09 \\
        Buckaroo & 0.0 & 0.02 & Build2 & 0.0 & 0.01\\
    \bottomrule
\end{tabular}
\end{table}
}{%
\begin{table}[t]
  \caption{The use of package management tools.}
  \saveSpaceFig
  \label{tab:popularity_pm}
  \scalebox{0.8}{
  \begin{tabular}{lcclcc}
    \toprule
        Install & Dep.(\%) & Repo.(\%) & Build & Dep(\%) & Repo(\%)\\
    \midrule
        Gitsubmod & 4.94 & 16.79& Make Only & 2.27 & 25.00 \\
        Deb & 9.32 & 5.90 & Cmake & 51.33 & 39.11 \\
        Pkg-config & 6.94 & 1.83 & MSBuild & 5.34 & 16.86 \\
        Conan & 0.45 & 0.99 & Autoconf & 12.3 & 15.27 \\
        Vcpkg & 0.56 & 0.39 & Meson & 3.0 & 2.11 \\
        Hunter & 0.18 & 0.2 & Bazel & 0.81 &1.37\\
        Clib & 0.05 & 0.12 & Buck & 0.0 & 0.27 \\
        Cpm & 0.07 & 0.1 & Xmake & 0.08 & 0.09 \\
        Buckaroo & 0.0 & 0.02 & Build2 & 0.0 & 0.01\\
    \bottomrule
\end{tabular}
}
\saveSpaceFig
\end{table}
}

Only 1.8\% of repositories (442/24,001, the sum of Conan, Vcpkg, Hunter, Clib, Cpm, and Buckaroo) adopt application-level package manager toolchains that introduce 1.3\% of dependencies, even if it provides the best capability to manage TPLs like \verb|Maven|. Between application-level package managers, we notice that Conan and Vcpkg contribute about 80\% of usage. Other tools are rarely used in real-world \ccpp projects. 
Interestingly, Conan and Vcpkg choose a different design philosophy from others. They directly use the ready-to-use binary packages rather than source code packages to construct their central repositories. This strategy eases the effort of users to compile the code on their own. We believe it is the future direction for \ccpp package managers.

We notice that traditional tools that do not have the capability to manage TPLs still have a high prevalence. Modern tools, including six application-level package managers mentioned above and two build systems (Xmake and Build2) that integrate a \ccpp package manager, need to extend their prevalence.




\aptLtoX[graphic=no,type=env]{%
\begin{framed}
Finding-2: \ding{172} For repositories that use formal methods to explicitly specify dependency installation, developers prefer to implement dependency installation through the naive code clone method. There is no convention at present to use a package manager in the \ccpp ecosystem.  \ding{173} Less than 2\% of repositories adopt application-level package managers, which means package managers are still in the primary stage in the \ccpp ecosystem.
\end{framed}
}{%
\vskip 2mm
\begin{boxedtext}
Finding-2: \ding{172} For repositories that use formal methods to explicitly specify dependency installation, developers prefer to implement dependency installation through the naive code clone method. There is no convention at present to use a package manager in the \ccpp ecosystem.  \ding{173} Less than 2\% of repositories adopt application-level package managers, which means package managers are still in the primary stage in the \ccpp ecosystem.
\end{boxedtext}
}

\subsubsection{\textbf{Combinations of reuse methods}} Since \ccpp dependency lifecycle is separated into two phases. Toolchains are formed by the combination of dependency installation tools and build systems. For each phase, there are many available tools. As a result, countless toolchains can be formed and used in \ccpp projects. We extract all combinations of tools in \ccrepodataset to investigate toolchains in the \ccpp ecosystem.

\begin{figure}[t]
    \centering
    \includegraphics[width=0.99\columnwidth]{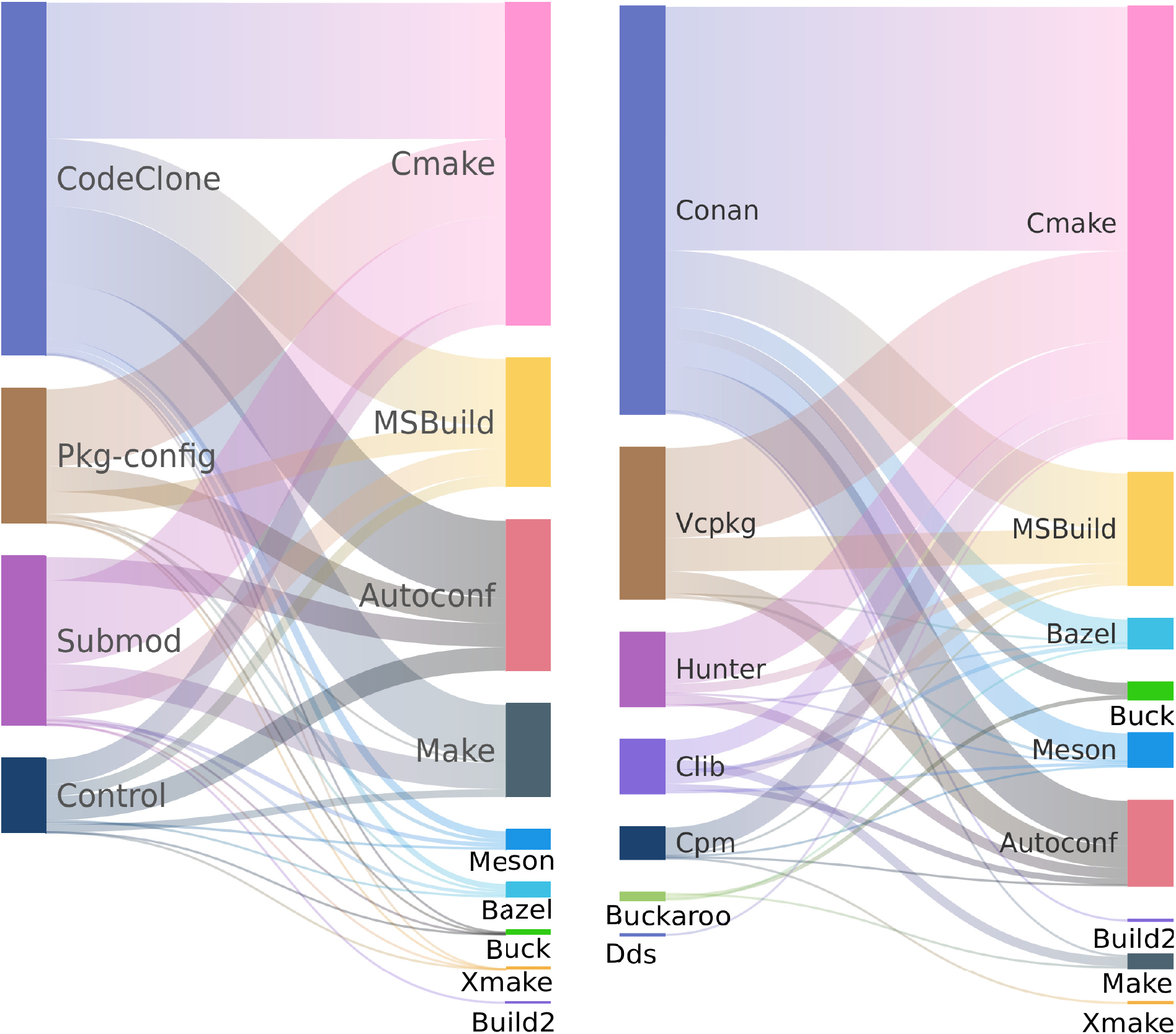}
    \saveSpaceFig
    \caption{Combinations of reuse methods in dependency lifecycle.}
    \saveSpaceFig
    \label{fig:toolchain_sankey}
\end{figure}

Results are shown in \Fig~\ref{fig:toolchain_sankey}. Two graphs are divided by traditional tools (shown on the left) and modern tools (shown on the left) in \phaseone phase. For the sake of typography, the sub-graph on the right is scaled up to the same size of the left sub-graph.
The figure shows that each tool in \phaseone phase can be combined with any tool in \phasetwo phase to form a toolchain.
It reveals a serious problem that TPL management methods in the \ccpp ecosystem are chaotic. There is no unified package management tool for \ccpp, moreover, the separation of \ccpp dependency lifecycle further increases the complexity and fragmentation of the usage of tools with an exponential growth rate. Chaotic usage of tools increases the development cost on library reuse, collaboration and integration.


For both traditional tools and modern tools in \phaseone phase, Cmake, MSBuild, and Autoconf are always the main options to build the software, which is consistent with \Tab~\ref{tab:popularity_pm}. Interestingly, we find that Make is a main choice for traditional tools in \phaseone phase, however, application-level package managers rarely work with Make. Furthermore, \Tab~\ref{tab:popularity_pm} shows that Make is too naive to handle dependencies. Despite the fact that it is the most important build system in \ccpp ecosystem and more than 80\% projects depend on it in \phasetwo phase, only 2.27\% of dependencies are added through it.

Even for modern tools in \phaseone phase, they rarely work with modern build systems. This is because package managers need to first consider integrating with mainstream build systems due to the separation of \ccpp dependency management. The separation impedes the formation of advanced toolchains. 

\aptLtoX[graphic=no,type=env]{%
\begin{framed}
Finding-3: \ding{172}Due to the lack of a unified package manager and the separation of TPL dependency lifecycle, the usage of toolchains is chaotic in the \ccpp ecosystem. \ding{173} Compatibility issues exist in toolchains for \ccpp. Make is a usual option for traditional TPL reuse methods, but it is not effective to work with modern application-level package managers. \ding{174} Basic build systems, including Make, Cmake, MSBuild, and Autoconf, still dominate the usage in the ecosystem despite the fact that many modern build systems have been proposed.
\end{framed}
}{%
\vskip 2mm
\begin{boxedtext}
Finding-3: \ding{172}Due to the lack of a unified package manager and the separation of TPL dependency lifecycle, the usage of toolchains is chaotic in the \ccpp ecosystem. \ding{173} Compatibility issues exist in toolchains for \ccpp. Make is a usual option for traditional TPL reuse methods, but it is not effective to work with modern application-level package managers. \ding{174} Basic build systems, including Make, Cmake, MSBuild, and Autoconf, still dominate the usage in the ecosystem despite the fact that many modern build systems have been proposed.
\end{boxedtext}
}

\subsection{RQ2: TPL Data Scope}\label{sec:data scope}
The goal of this section is to understand the TPL data scope in the \ccpp ecosystem. As described in \Sec~\ref{sec:dependency_cycle}, there are three categories of TPL databases, including system mirrors, central repositories of package managers, and source code hosts. For source code hosts, GitHub repositories are often collected as the local TPL database for detection~\cite{duan2017identifying, ban2021b2smatcher, woo2021centris}. However, it is not clear which database is better and whether libraries in databases are sufficient and high-quality. Therefore, we conduct a comparative analysis on three categories of databases to answer two questions in the following two subsections: (1) what is the data coverage of each TPL database? (2) is it reasonable to use GitHub repositories as the local TPL database like what related work does?

\subsubsection{\textbf{Database Coverage.}}\label{sec:database coverate}
\begin{figure}[t]
    \centering
    \includegraphics[width=0.77\columnwidth]{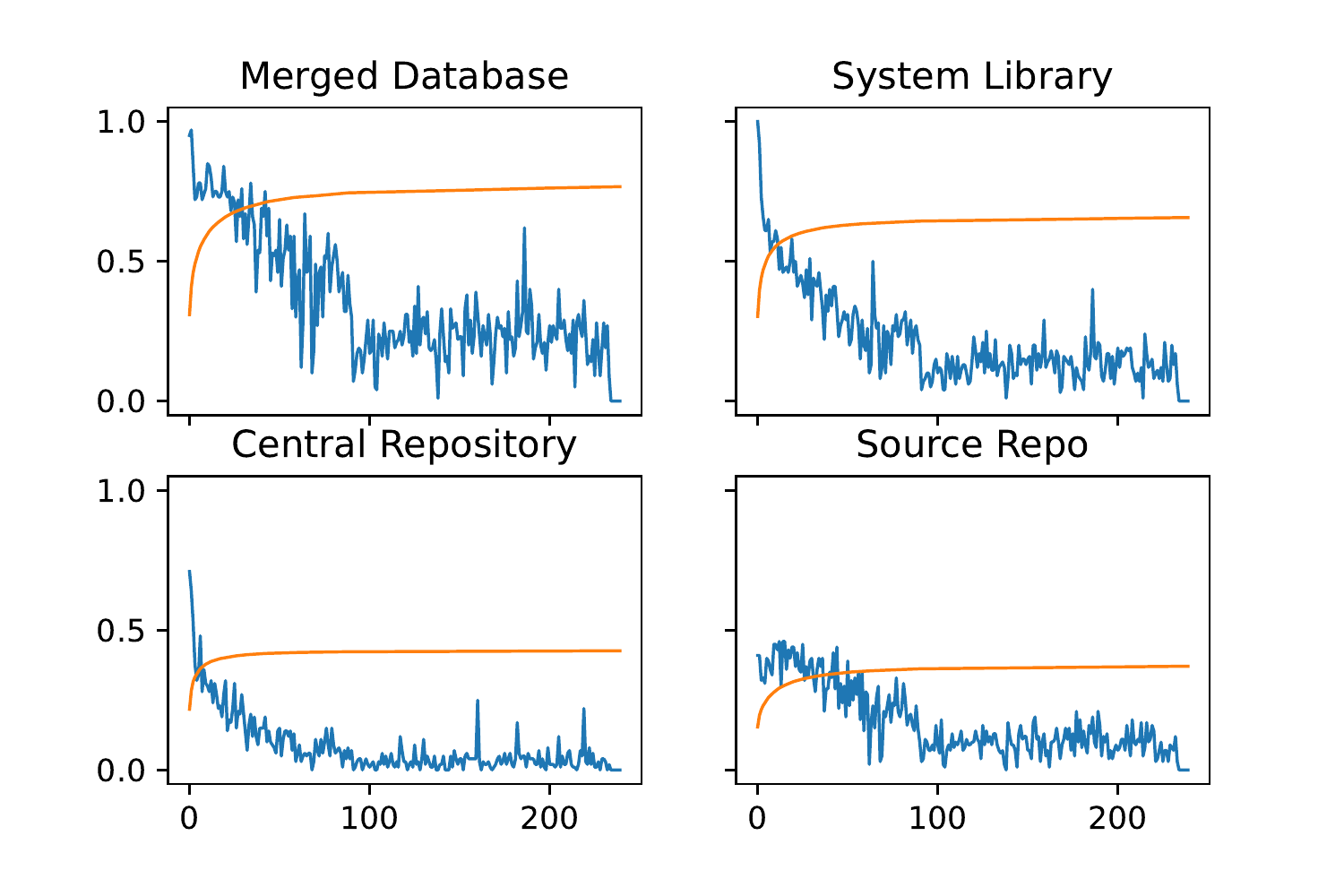}
    \saveSpaceFig
    \caption{TPL data coverage of all database. The blue line represents data coverage of every 100 libraries. Yellow line indicates the accumulated number of dependencies that are contributed by covered libraries.}
    \saveSpaceFig
    \label{fig:database_coverage}
\end{figure}

It is important for TPL database to have a large coverage since it defines the availability and effectiveness of package managers in practice. Moreover, it is beneficial for the improvement of the recall rate of detection as mentioned in \Sec~\ref{existing_sca_tools}
From extracted 150K dependencies, we have identified 23,322 reused libraries. \Fig~\ref{fig:database_coverage} presents the data coverage of the three individual databases and a merged database that contains all libraries of three databases. We sort all reused libraries by their popularity (i.e. the number of reuses) and generate a batch every 100 libraries on the horizontal axis. The data coverage indicated by blue lines is calculated by counting how many libraries of a batch exist in the database. For example, the first point (0, 1) in the sub-figure of system library means that all top 100 popular libraries exist in the system library database.

From \Fig~\ref{fig:database_coverage} we can see that system library database has the largest coverage on all reused libraries and it contains the largest number of popular libraries. All top 100 popular libraries and 5,163 reused libraries can be found in the system library database that covers a total number of 65.5\% of dependencies (98,896). Central repositories of application-level package managers have a lower coverage (70\%) on top 100 popular libraries than system library database, but results show that 58\% of libraries in central repositories are top 20\% popular libraries which means these central repositories have high-quality libraries that may be commonly used by developers. In contrast, more than 60\% of libraries in the other two databases are not detected as reused libraries. For source code repositories, the database of \centris has the lowest coverage on popular libraries that drops to 41\%\~44\% and the coverage on top 100 popular libraries is less than 50\%. From the merged database, we notice that the three categories of databases cover 77\% of dependencies with 8,505 reused libraries.

Even though the system library database has the largest data coverage, it only contains 22\% of reused libraries. The merged database of three categories contains 36\% of reused libraries. After eliminating overlapped libraries between three categories of databases, each one still has more than 50\%, 73\% and 98\% left as unique libraries, respectively for central repository database, GitHub repositories, and system library database. It suggests that TPLs in \ccpp ecosystem are scattered on the web. There is no complete TPL database for the ecosystem. 

Through checking overlapped libraries, we find that data fragmentation causes a problem that overlapped libraries are not consistent with each other. For example, Debian maintains LibPNG which is downstream of original official repository. New versions are created to fix vulnerabilities by Debian team and a Debian revision version is attached following original semantic version. This causes differences in terms of versioning and security information. Version number would be ineffective to report potential vulnerabilities. For instance, CVE 2019-7317 exists in libpng 1.6.x before 1.6.37, however, it does not affect Debian libpng after 1.6.36-4. Moreover, there are two forms of packages for a \ccpp library, source code repository and binary package. This kind of difference would likewise affect the vulnerability reporting in security management since a source repository might be compiled into multiple binary packages. If a vulnerability is bundled with a source repository, it may not be contained in some binary packages. For example, CVE-2021-38171 only affects the libavformat package in FFmpeg, we scan the dependencies in Debian mirrors and find 112 repositories depend on FFmpeg, however, 12/112 repositories do not depend on the vulnerable libavformat package.

\aptLtoX[graphic=no,type=env]{%
\begin{framed}
Finding-4: \ding{172} There is no unified and complete TPL database in the \ccpp ecosystem. Each category of database has more unique libraries than overlapped libraries. The merged database only contains a small portion of reused libraries (36\%).
\ding{173} TPL data fragmentation exists in \ccpp ecosystem. Libraries are inconsistent between databases, that may threaten the effectiveness of TPL detection and vulnerability reporting.
\end{framed}
}{%
\vskip 2mm
\begin{boxedtext}
Finding-4: \ding{172} There is no unified and complete TPL database in the \ccpp ecosystem. Each category of database has more unique libraries than overlapped libraries. The merged database only contains a small portion of reused libraries (36\%).
\ding{173} TPL data fragmentation exists in \ccpp ecosystem. Libraries are inconsistent between databases, that may threaten the effectiveness of TPL detection and vulnerability reporting.
\end{boxedtext}
}




\subsubsection{\textbf{GitHub TPL Coverage.}}\label{sec:github coverate} Nowadays, GitHub is the main platform to host source code repositories. Researchers naturally consider collecting GitHub repositories as a TPL database. The mainstream solution to local TPL database construction in related work is to collect GitHub repositories with a significant stargazer count~\cite{duan2017identifying, ban2021b2smatcher, woo2021centris}. However, \ccpp is an older programming language than GitHub. There are a large number of \ccpp libraries that are hosted in Linux distributions, SourceForge, etc. Therefore, it is questionable whether GitHub repositories have convincing representativeness as TPL database. Collecting GitHub repositories to build TPL database is based on an assumption that popular repositories on GitHub are likely to be reused as TPLs in software development. There are two serious problems with this assumption: (1) it assumes that most libraries are hosted on GitHub. (2) repositories on GitHub are reusable libraries.

To evaluate the first assumption, we use the collection of libraries in central repositories as the ground truth to figure out how many libraries are hosted on GitHub. Since application-level package managers are designed for TPL reuse management, packages in their central repositories can be treated as TPLs. To keep consistent with previous works~\cite{duan2017identifying, woo2021centris}, we divide all source code repositories into three categories, GitHub repositories with more than 100 stars, GitHub repositories with less than 100 stars, and non-GitHub repositories.

\begin{figure}[t]
    \centering
    \includegraphics[width=0.77\columnwidth]{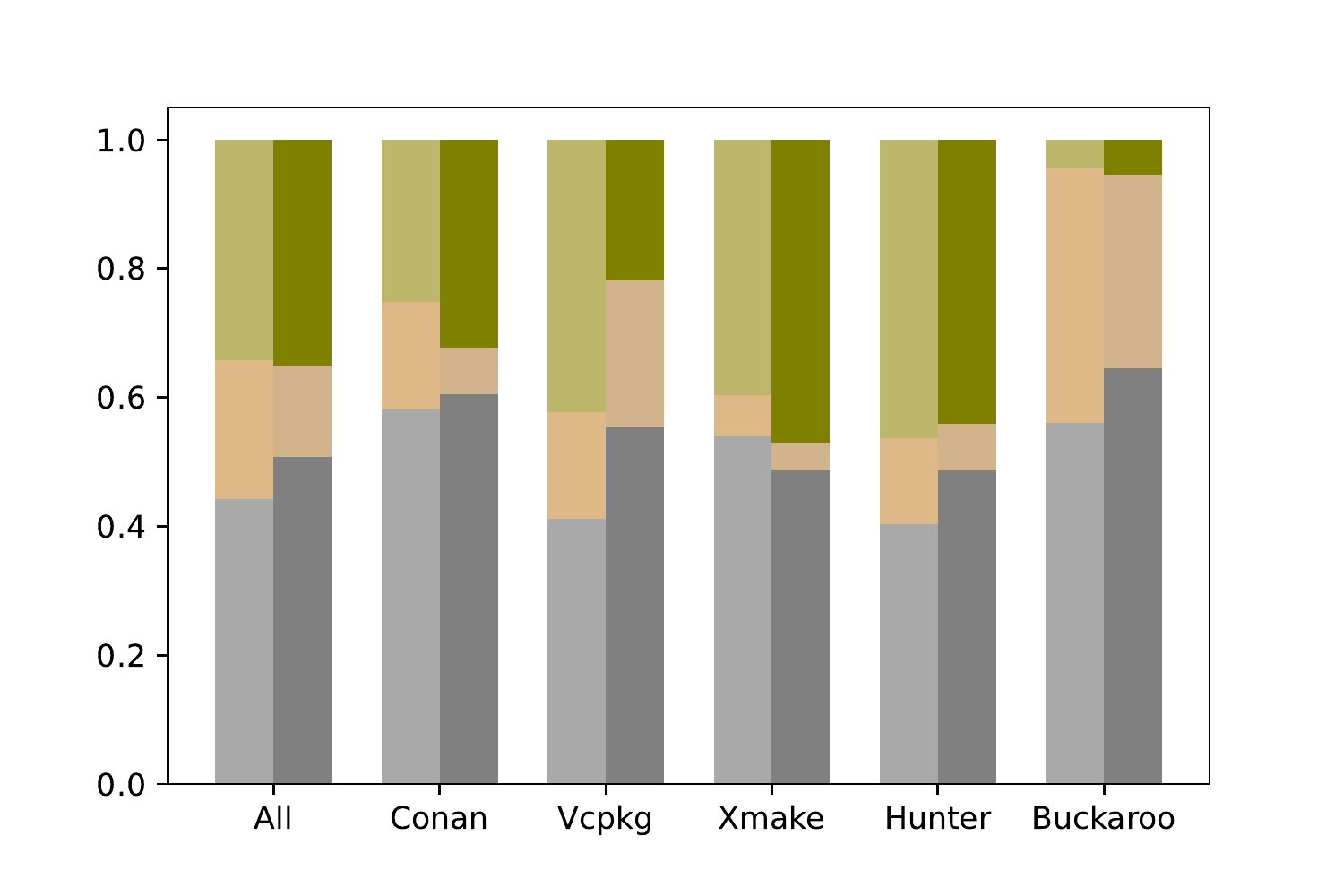}
    \saveSpaceFig
    \caption{Distribution of data sources for overall libraries (left) and reused libraries (right) for each package manager. The bottom layer represents GitHub repositories with more than 100 stars, the middle layer: less than 100 stars, the top layer: not GitHub repositories.}
    \saveSpaceFig
    \label{fig:distribution_reused_libs}
\end{figure}

\Fig~\ref{fig:distribution_reused_libs} shows the distribution of data sources for overall libraries and reused libraries for each package manager. We can see that overall, 43\% of libraries and 50\% of reused libraries are hosted on GitHub with more than 100 stars. It means that collecting repositories with more than 100 stars will miss half of libraries. About 20\% of libraries are GitHub repositories with less than 100 stars. However, we do not recommend practitioners to crawl all repositories with few stars, since a lot of noise data would be collected due to the long-tail distribution of stars. On GitHub, 41K \ccpp repositories have more than 50 stars and more than 2 million non-folk \ccpp repositories are hosted on GitHub.

To evaluate the second assumption, we inspected the false positives mentioned in \Sec~\ref{existing_sca_tools}.
We noticed that there are a significant number of repositories that are not TPLs and would not be reused by other developers. 
Such repositories are the main reason for false positives in results of \centris. We propose two concepts for classifying source repositories, application and library. An application is a computer program that is designed for specific users. A library is a collection of code to make other programs work. They provide special features to developers. The most remarkable difference is that applications are user-oriented, and libraries are developer-oriented. It is crucial to distinguish applications and libraries to build a TPL database. 

\aptLtoX[graphic=no,type=env]{%
\begin{framed}
Finding-5: Collecting GitHub repositories with more than 100 stars is not a reasonable approach to building TPL database. Only half of libraries of \ccpp package managers could be collected in this way. Moreover, it would collect a significant number of applications that are not TPLs, leading to a low precision (33.6\%) of \centris for TPL detection.
\end{framed}
}{%
\vskip 2mm
\begin{boxedtext}
Finding-5: Collecting GitHub repositories with more than 100 stars is not a reasonable approach to building TPL database. Only half of libraries of \ccpp package managers could be collected in this way. Moreover, it would collect a significant number of applications that are not TPLs, leading to a low precision (33.6\%) of \centris for TPL detection.
\end{boxedtext}
}


\subsection{RQ3: Key Libraries to \ccpp Ecosystem}
A library is considered more popular if it is reused by more repositories. Key libraries have a considerable impact on the whole ecosystem. It is a critical research question in previous empirical studies for other ecosystems~\cite{wang2020empirical, liu2022demystifying, he2021large}. We conduct an analysis on the popularity of libraries in the \ccpp ecosystem to answer questions: (1) what is the library popularity distribution? (2) what are the commonly reused libraries and how do they affect the ecosystem?

\subsubsection{\textbf{Library popularity distribution}}
A total number of 150,927 TPL dependencies are extracted from \ccrepodataset. On average, each \ccpp repository depends on 6.3 external libraries. As for the library usage intensity, 42.4 \% (10179 / 24001) of repositories do not depend on TPLs and 41.4\% of repositories (9,934) adopt at most 10 libraries. Only 2.1 \% of repositories (502) reuse more than 50 libraries. 23,322 distinct libraries are identified and each library is reused 6.5 times on average. The median of reuse times is 1. It means that a highly polarized popularity structure exists in \ccpp libraries. Top 1\%, 5\%, 10\% and 20\% of libraries contribute 45\%, 68\%, 77\% and 84\% of dependencies, respectively. The popularity distribution highly meets the pareto distribution that describes social behaviors, i.e., most dependencies are contributed by a small group of libraries. In economics, the Gini coefficient is a measure of statistical dispersion intended to describe the income polarization. We calculate the Gini coefficient of popularity as 0.79 that is extremely unequal distribution in human society.

We scan \ccrepodataset over the last ten years and find that the polarized popularity structure has intensified over time. In 2011, the Gini coefficient of popularity is 0.69 and Top 1\%, 5\%, 10\% and 20\% of libraries contribute 29\%, 54\%, 65\% and 76\% of dependencies, respectively. 

A highly polarized popularity structure is a double-edged sword. On one hand, the majority of the dependencies can be covered with a small TPL database. It would be effective and efficient to build such a small database for library management and dependency detection. Data collection becomes much easier and less time-consuming. On the other hand, the \ccpp ecosystem heavily depends on a small group of critical libraries that may threaten the security of the entire ecosystem. If a library contains a vulnerability with high severity, it would be a disaster to the ecosystem such as the spread of the Heartbleed Vulnerability~\cite{durumeric2014matter} in the most popular communication TPL, OpenSSL.

\aptLtoX[graphic=no,type=env]{%
\begin{framed}
Finding-6: A highly polarized popularity structure exists in \ccpp TPLs and has intensified over the last ten years. Practitioners in areas of TPL management and detection could focus on a small group of TPL data. However, the highly polarized popularity structure may threaten the security of the entire ecosystem.
\end{framed}
}{%
\vskip 2mm
\begin{boxedtext}
Finding-6: A highly polarized popularity structure exists in \ccpp TPLs and has intensified over the last ten years. Practitioners in areas of TPL management and detection could focus on a small group of TPL data. However, the highly polarized popularity structure may threaten the security of the entire ecosystem.
\end{boxedtext}
}

\subsubsection{\textbf{Popular TPL Influence}}
To explore the extent to which popular libraries affect the whole ecosystem, we select the top 100 libraries to observe the direct and transitive dependencies. Through checking all direct dependencies, we notice that some repositories are reused as libraries by other repositories. Therefore, we can generate the dependency chain between repositories to identify transitive dependencies. We find that it is a big research challenge to construct a whole dependency graph for the \ccpp ecosystem. More details are described in \Sec~\ref{sec:limitation}. Therefore, we only resolve the transitive dependencies for top 100 libraries.
\Tab~\ref{tab:top10libs} shows top 10 popular libraries from direct dependencies and transitive dependencies. We ignore compile-time dependencies such as pkg-config and test-time dependencies like googletest since they are not required in the runtime phase.

\aptLtoX[graphic=no,type=env]{%
\begin{table}[t]
  \caption{Top 10 popular libraries. STD Lib means standard library.}
  \label{tab:top10libs}
  \begin{tabular}{llrlll}
    \toprule
        Direct & Type & Repo & Transitive & Type & Repo\\
    \midrule
        Threads & STD Lib     & 2825&   Threads & STD Lib     &6459\\
        Zlib    & Compression & 1977&   Zlib    & Compression &5212\\
        Boost   & Programming & 1676&   Libm    & STD Lib     &5124\\
        OpenCV  & CV          &  987&   Libdl   & STD Lib     &3830\\
        OpenSSL & Crypto      &  929&   OpenSSL & Crypto      &3387\\
        Libm    & STD Lib     &  890&   Cuda    & DEV ENV     &3291\\
        OpenGL  & Graphics    &  876&   Librt   & STD Lib     &3191\\
        Eigen   & Math        &  776&   OpenMP  & Programming &2999\\
        OpenMP  & Programming &  774&   Libsocket & Network   &2971\\
        X11     & GUI         &  737&   Libnsl  & Network     &2934\\
    \bottomrule
\end{tabular}
\end{table}
}{%
\begin{table}[t]
  \caption{Top 10 popular libraries. STD Lib means standard library.}
  \saveSpaceFig
  \label{tab:top10libs}
  \scalebox{0.7}{
  \begin{tabular}{llrlll}
    \toprule
        Direct & Type & Repo & Transitive & Type & Repo\\
    \midrule
        Threads & STD Lib     & 2825&   Threads & STD Lib     &6459\\
        Zlib    & Compression & 1977&   Zlib    & Compression &5212\\
        Boost   & Programming & 1676&   Libm    & STD Lib     &5124\\
        OpenCV  & CV          &  987&   Libdl   & STD Lib     &3830\\
        OpenSSL & Crypto      &  929&   OpenSSL & Crypto      &3387\\
        Libm    & STD Lib     &  890&   Cuda    & DEV ENV     &3291\\
        OpenGL  & Graphics    &  876&   Librt   & STD Lib     &3191\\
        Eigen   & Math        &  776&   OpenMP  & Programming &2999\\
        OpenMP  & Programming &  774&   Libsocket & Network   &2971\\
        X11     & GUI         &  737&   Libnsl  & Network     &2934\\
    \bottomrule
\end{tabular}
}
\saveSpaceFig
\end{table}
}

Considering direct dependencies, a set of top 10 popular libraries affect 28\% of repositories (6754) and the top 100 libraries affect 42\% (10100). Since 13,822 repositories have dependencies, the top 100 popular libraries affect 73\% of repositories with dependencies. For transitive dependencies, the top 10 popular libraries have an impact on the entire ecosystem.

Moreover, compared to the results of previous studies, we notice that a large number of dependencies of system libraries are ignored by previous studies. However, our statistical results show that system libraries are the most important category of libraries.
They occupy a large portion of dependencies, especially transitive dependencies. All top 10 popular libraries of transitive dependencies are low-level system libraries that are default installed on OS. 
Previous empirical studies investigate \ccpp dependencies using clone-based detection techniques. As a consequence, only TPLs contained in detection targets can be detected. It would reduce the number of libraries discovered. The libraries (e.g. system libraries) that are not used by code clone would be missing. Since \ccpp are low-level programming languages, a large number of \ccpp libraries are installed on OS by default, that are downloaded through system-level package managers and installed globally. If an application relies on system libraries, build tools can directly locate the reused libraries under system paths without copying the code of libraries. Therefore, the previous studies lack the capability to extract dependencies of system libraries and all of their findings are explored around non-system libraries.


\aptLtoX[graphic=no,type=env]{%
\begin{framed}
Finding-7: System libraries on OS are the most important libraries for the \ccpp ecosystem. A group of 10 popular system libraries have an impact on the entire ecosystem. However, system libraries are always neglected in TPL management and detection.
\end{framed}
}{%
\vskip 2mm
\begin{boxedtext}
Finding-7: System libraries on OS are the most important libraries for the \ccpp ecosystem. A group of 10 popular system libraries have an impact on the entire ecosystem. However, system libraries are always neglected in TPL management and detection.
\end{boxedtext}
}


\subsection{RQ4: TPL Version Selection and Impact }\label{sec:security}
Version selection is an important behavior in TPL reuse. It could directly affect library update and vulnerability propagation~\cite{huang2020interactive, liu2022demystifying}. To demystify TPL version selection and its impact in the \ccpp ecosystem, we try to answer two questions in the following two subsections respectively: (1) do developers specify version constraints for dependencies? do they update them? 2) How do vulnerabilities in TPLs affect the \ccpp ecosystem?

\subsubsection{\textbf{TPL Version Management}}
Version specification indicates the compatibility of dependencies and well-maintained versioning constraints can bring many benefits, such as avoiding using vulnerable versions. Many works~\cite{10.1145/3372297.3417232, jafari2021dependency, chinthanet2021lags, prana2021out, gkortzis2021software, zimmermann2019small, liu2022demystifying} investigate the security landscape based on version specifications in other language ecosystems. However, specifying version constraints is not always a requirement and there are few relevant regulations and conventions for \ccpp. To investigate version constraints in the \ccpp ecosystem, \tool parses statements about dependency version for every package management tool to extract version specifications. 

In total, we obtain 35,349 dependencies that specify versions. 
Unlike other modern languages that embrace semantic versioning to specify their required TPL versions, dependency declaration in \ccpp projects has a quite low ratio of explicitly specifying the versions of dependencies (27\%). Such a low ratio is an indication of inadequate control of version constraints in the \ccpp ecosystem. To understand the reasons, we examine the stages in dependency lifecycle that introduce the TPLs without version specification. 
We find that 51.7 \% of dependencies handled by tools during \phaseone phase specify versions. However, the number drops to only 8.09\% in the \phasetwo phase. Furthermore, for package managers that contain multi-version in their central repository like Conan, 97\% of dependencies specify versions. For package managers that do not support multiple versions for one package, like system-level package managers, only 18.4\% specify versions. Moreover, we notice that no low-level system libraries specify version constraints in their dependencies. The OS environment takes responsibility for maintaining system libraries, such as patching and updating. That explains why the system package managers have a low ratio of version constraints. It is not necessary to consider the problem of library versions once the OS environment is determined. As for updates, we select the top 10 popular libraries with version constraints and calculate how many dependencies adopt the latest versions. Results show that only 9.5\% of constraints specify the latest versions. 

\aptLtoX[graphic=no,type=env]{%
\begin{framed}
Finding-8: \ding{172} Version specifications are more widespread in \phaseone phase, especially for package managers that support multiple versions. \ding{173} The versions of system libraries are determined by OS. Therefore, dependencies that have requirements on system library versions might be not compatible with some systems.
\end{framed}
}{%
\vskip 2mm
\begin{boxedtext}
Finding-8: \ding{172} Version specifications are more widespread in \phaseone phase, especially for package managers that support multiple versions. \ding{173} The versions of system libraries are determined by OS. Therefore, dependencies that have requirements on system library versions might be not compatible with some systems.
\end{boxedtext}
}

\subsubsection{\textbf{Vulnerable TPL Influence}}
An important version-based analysis is to investigate how vulnerable libraries affect the language ecosystem~\cite{gkortzis2021software, decan2018impact, zimmermann2019small, zerouali2021impact, liu2022demystifying}. Different from other language ecosystems where the allowed versions are usually explicitly specified and vulnerabilities can be easily tracked, only a small part of dependencies are declared with specified versions in \ccpp projects so that \phasetwo phase could find the default version for system libraries on OS. To study the impact of vulnerable libraries, we use the latest Debian security information to match dependencies without version constraints, supposing that these dependencies reuse the latest versions of libraries on OS. We have got 6661 unpatched vulnerabilities existing in \ccpp repositories on Debian. As for dependencies with version specifications, we match specified versions to a vulnerability database of a commercial security company that contains the vulnerability list of 299 reused libraries.

For dependencies without version specifications, 13\% of dependencies (13463) reuse vulnerable libraries involving 22\% of repositories (5315) through direct dependencies. 
Moreover, the affection rate will become much higher if the transitive dependencies are taken into consideration.
By checking management tools related to vulnerable dependencies, we find that all vulnerable dependencies come from build systems and management tools in Linux ecosystem like Deb, i.e., the system library toolchain. It relies heavily on OS environment and the system libraries. An advantage is that security issues would be fixed by OS maintainers. It lightens and simplifies the tasks of developers.

For dependencies with version specifications given by developers, 1143 dependencies are found to use vulnerable versions involving 118 reused libraries and 758 upstream dependent repositories. We find that 49.3\% of dependencies with version specifications libraries use vulnerable versions. It means that poorly maintained versions cause a significant security risk to dependent applications. We also find that it is challenging to recall introduced vulnerabilities based on library name and version for \ccpp.

\aptLtoX[graphic=no,type=env]{%
\begin{framed}
Finding-9: Management tools that rely on system libraries will introduce vulnerabilities from OS environments. Developers would not explicitly specify versions using these tools and the OS maintainer is responsible for eliminating vulnerabilities. About half of dependencies with version specifications of vulnerable libraries use vulnerable versions, which requires developers to update the dependencies manually.
\end{framed}
}{%
\vskip 2mm
\begin{boxedtext}
Finding-9: Management tools that rely on system libraries will introduce vulnerabilities from OS environments. Developers would not explicitly specify versions using these tools and the OS maintainer is responsible for eliminating vulnerabilities. About half of dependencies with version specifications of vulnerable libraries use vulnerable versions, which requires developers to update the dependencies manually.
\end{boxedtext}
}











\section{Discussion}\label{sec:fw}

\subsection{Implications}
\noindent \textbf{For \ccpp Developers.} We recommend \ccpp developers to use application-level package managers in the development and avoid using customized documents to describe dependency installation. Application-level package managers require developers to facilitate the use of dynamic linking rather than code clone. Due to the small size of central repositories, reused libraries may be not hosted in public official repositories of package managers. It poses challenges and difficulties for the use of package managers in practice. To address this issue, developers could create a package repository to store TPLs for public release and join the ecosystem construction for \ccpp. Developers should be aware of implicit OS library dependencies and convert them to explicit ones through management tools while it would require more effort. When code clone, especially partial clone, appears, package managers do not apply to cloned code. It is suggested to adopt a SBOM mechanism to manage cloned code if code clone is unavoidable.


\noindent \textbf{For \ccpp TPL Auditors.} Due to the lack of a unified package manager and the features of \ccpp, it is more complicated for \ccpp to detect reused TPLs than other languages. Existing detection tools only have a limited capability to resolve and track the whole dependency lifecycle. Our findings help practitioners to develop more effective TPL detection systems for \ccpp and build TPL databases more effectively. We conclude some recommendations for \ccpp TPL Auditors: \ding{172} parsing build scripts in \phasetwo to recall more dependencies; \ding{173} scanning OS environment for system libraries; \ding{174} collecting TPL data from three categories of databases; \ding{175} distinguishing applications and libraries for source code repositories; \ding{176} collecting a database from popular libraries if resources are limited; \ding{177} focusing on differences caused by data fragmentation.

\noindent \textbf{For \ccpp Package Managers Designers} 
Existing system package managers are not designed for \ccpp package management and lack some advanced features. As shown in \Fig~\ref{fig:dependency_lifecycle}, system libraries are installed through commands like \verb|apt install| and imported directly by build systems. Designers should add a SBOM file to describe dependencies and reproduce the environment automatically. \verb|Dockerfile| in docker is a good practice, but there is no such mechanism for system package managers on Linux. Many developers rely on system package managers, even they are not designed for development. Since system package managers only provide one version for each package, they do not support evolving needs of developers. Besides, they do not provide security alert when users import a vulnerable package. Designers could consider to provide multiple versions and track introduced vulnerabilities to support effective and secure development.
For application-level package managers, it is almost impossible to build a central and trusted database that contains sufficient \ccpp libraries. We recommend designers should consider making application-level package managers compatible with existing library databases, such as Debian mirrors, that have the largest coverage and massive packages. Due to the separation of dependency lifecycle and chaotic combinations of tools, package managers should integrate with all mainstream build systems to complete the whole toolchain automatically and become more ease-of-use.

\subsection{Limitations}\label{sec:limitation}
\noindent \textbf{Code clone detection} In our work, we integrate unchanged \centris for code clone detection. \centris is a state-of-the-art tool to detect \ccpp dependencies. However, it is not precise or efficient for large-scale empirical studies as mentioned in \Sec~\ref{accuracy}. It makes our empirical study lack dependencies that are caused by code clone in the evolutionary analysis.

\noindent \textbf{Dependency chain} 
Macros and compilation settings change dependencies of a project. It is decided in the period of compilation and cannot be identified by static code analysis. For example, there are dozens of flags in the build scripts of FFmpeg~\cite{url:ffmpeg}. Each flag can be set to enabled to control whether a TPL needs to be introduced during the \phasetwo phase. Besides, as mentioned in \Sec~\ref{sec:database coverate}, dependencies on source code repositories may not keep the same with binary packages after compilation. It also affects the dependency chain construction.



\section{related work}\label{sec:relwork}


\noindent \textbf{Dependency Management \& Detection.}
To properly manage dependencies of user projects, lots of research has investigated the way how dependencies are integrated and the accompanied problems.

For C/C++, Miranda et al.~\cite{miranda2018use} performed a questionnaire survey and collected opinions on the use of package managers by 343 C++ developers from 42 open-source projects. Centris~\cite{woo2021centris} and D{\'e}j{\`a}Vu~\cite{lopes2017dejavu} and the commercial product \fossid detect \ccpp TPL dependencies through code clone detection. OSSPolice~\cite{duan2017identifying} and Modx~\cite{yang2022modx} detect TPL dependencies in binaries. Besides, there are some famous SBOM scan tools, like OWASP~\cite{owasp} and \sonatype~\cite{sonatype}.

Besides, there are also many work investigating the dependency management in other languages. Dietrich et al.~\cite{dietrich2019dependency} studied the choices developers made on defining dependencies in 17 different package managers. BuildMedic~\cite{macho2018automatically}, Riddle~\cite{8812128}, and Sensor~\cite{9350237} study on build issues in Maven. Wang et al.~\cite{wang2018dependency, 8812128, 9350237, 10.1145/3377811.3380426, 9401974, li2022nufix} had done many interesting work to investigate dependency management issues in different languages.

\noindent \textbf{Dependency based Ecosystem Analysis.}
For \ccpp, there is few work, especially large-scale studies due to the lack of unified \ccpp package manager. Wu et al.~\cite{wu2015developers} analyzed 30 applications to discuss the usage of C++ standard libraries. OSSPolice~\cite{duan2017identifying} conducts a large-scale usage analysis in Apks.

Many researchers analyzed dependencies for other language ecosystems. Huang et al.~\cite{huang2022characterizing} and He et al.~\cite{he2021large} conducted studies on usages and migrations of libraries for Java. Zimmermann et al.~\cite{zimmermann2019small}, Decan et al.~\cite{decan2018impact}, Chengwei et al.~\cite{liu2022demystifying} investigate on security risks of dependencies for JavaScript. However, all of these vulnerability impact analysis are based on clear dependency relations, and our work can provide dependency relations for C/C++ projects in a confident manner, therefore can inspire more in-depth analyses on vulnerabilities in the ecosystem of C/C++.

\section{Conclusion}\label{sec:conclu}
In this paper, we first undertake an extensive investigation on how TPL dependencies are handled in \ccpp projects and summarize the lifecycle for \ccpp TPL dependencies. Based on the dependency lifecycle, we propose a comprehensive and precise \ccpp dependency detector, \tool. Experiments demonstrate that \tool is capable of scanning large-scale \ccpp repositories for empirical studies. We apply \ccpp to scan 24K Github repositories to conduct a large-scale empirical study on dependencies in \ccpp ecosystem. Our study unveils a lot of findings regarding TPL reuse method, TPL data, key TPLs and TPL version selection.

\begin{acks}
This work is supported by the Key Research Program of the Ministry of Science and Technology of China (no.2018YFF0215901), Nanyang Technological University (NTU)-DESAY SV Research Program under Grant 2018-0980, Singapore Ministry of Education (MOE) Academic Research Funding (AcRF) Tier 2 under Grant MOE-T2EP20120-0004, and the program of China Scholarships Council award (202006210393).
\end{acks}


\balance
\bibliographystyle{ACM-Reference-Format}




\balance

\end{document}